\documentclass{ws-procs961x669}            
\begin{document}
\title{E$_6$ inspired SUSY models with Custodial Symmetry}

\author{R. Nevzorov$^*$}

\address{Alikhanov Institute for Theoretical and Experimental Physics, Moscow, 117218, Russia\\
$^*$E-mail: nevzorov@itep.ru}

\begin{abstract}
The breakdown of $E_6$ within the supersymmetric (SUSY) Grand Unified Theories (GUTs)
can result in SUSY extensions of the standard model (SM) based on the SM gauge
group together with extra $U(1)$ gauge symmetry under which right--handed
neutrinos have zero charge. In these $U(1)_N$ extensions of the minimal
supersymmetric standard model (MSSM) a single discrete $\tilde{Z}^{H}_2$ symmetry
may be used to suppress the most dangerous operators, that give rise to proton
decay as well as non--diagonal flavour transitions at low energies. The SUSY models
under consideration involves $Z'$ and extra exotic matter beyond the MSSM.
We discuss leptogenesis within this SUSY model and argue that the extra exotic states
may lead to the non--standard Higgs decays.
\end{abstract}


\bodymatter

\section{Introduction}\label{aba:sec1}
Supersymmetric (SUSY) extensions of the standard model (SM)
allows one to embed SM into Grand Unified Theories (GUTs)
based on simple gauge groups such as $SU(5)$, $SO(10)$ or $E_6$.
Indeed, it was found that the electroweak (EW) and strong gauge
couplings extrapolated to high energies using the renormalisation
group equation (RGE) evolution converge to a common value at
some high energy scale in the framework of the minimal SUSY standard
model (MSSM) \cite{Ellis:1990wk,Langacker:1991an,Amaldi:1991cn,Anselmo:1991uu}.
The incorporation of the SM gauge interactions within GUTs permits,
in particular, to explain the peculiar assignment of $U(1)_Y$ charges
postulated in the SM.

It is well known that each family of quarks and leptons fills
in complete 16 dimensional spinor representation of $SO(10)$
that also predicts the existence of right--handed neutrino,
allowing it to be used for both the see--saw mechanism and
leptogenesis. In $N=1$ SUSY GUT based on $E_6$ the complete
fundamental $27$ representation, that decomposes under
$SO(10) \times U(1)_{\psi}$ subgroup as
\begin{equation}
27 \to \left(16,\,\frac{1}{\sqrt{24}}\right) \oplus
\left(10,\,-\frac{2}{\sqrt{24}}\right) \oplus
\left(1,\,\frac{4}{\sqrt{24}}\right)\,,
\label{1}
\end{equation}
contains Higgs doublet. It is assigned to
$\left(10,\,-\frac{2}{\sqrt{24}}\right)$.
The SM gauge bosons belong to the adjoint representation of
$E_6$, i.e. $78$--plet. In $N=2$ SUSY GUT based on the $E_8$ gauge group
all SM particles belong to $248$ dimensional representation of
$E_8$ that decomposes under its $E_6$ subgroup as follows
\begin{equation}
248 \to 78 \oplus\, 3\times 27\, \oplus\, 3\times \overline{27}\, \oplus \, 8\times 1\,.
\label{2}
\end{equation}

The local version of SUSY (supergravity) results in a partial unification
of the SM gauge interactions with gravity. However supergravity (SUGRA) is a
non--renormalisable theory and has to be considered as an effective low energy
limit of some renormalisable or even finite theory. Currently, the best candidate
for such an underlying theory, i.e. hypothetical single framework that explains
and links together all physical aspects of the universe, is ten--dimensional
heterotic superstring theory based on $E_8\times E'_8$ \cite{5}.
Compactification of extra dimensions leads to an effective supergravity
and results in the breakdown of $E_8$ to $E_6$ or its subgroups in the
observable sector. The remaining $E'_8$ plays the role of a hidden sector
which gives rise to spontaneous breakdown of SUGRA.

\section{The $U(1)_N$ extensions of the MSSM}
In orbifold SUSY GUTs the $E_6$ gauge group can be broken down to
$SU(3)_{C}\times SU(2)_{W}\times U(1)_{Y}\times U(1)_{\chi}\times U(1)_{\psi}$,
where the $U(1)_{\psi}$ and $U(1)_{\chi}$ symmetries are defined by:
$E_6\to SO(10)\times U(1)_{\psi}$, $SO(10)\to SU(5)\times U(1)_{\chi}$ \cite{Nevzorov:2012hs}.
In order to ensure anomaly cancellation in this case the particle content
below the GUT scale $M_X$ should be extended to include three $27$--plets.
Each $27$--plet, referred to as $27_i$ with $i=1,2,3$, includes one generation
of ordinary matter, a SM singlet field $S_i$ (see last term in Eq.~(\ref{1})),
that carries non--zero $U(1)_{\psi}$ charge, as well as Higgs--like doublets
($H^{u}_{i}$ and $H^{d}_{i}$) and charged $\pm 1/3$ exotic quarks ($D_i$ and
$\bar{D}_i$) which are associated with $\left(10,\,-\frac{2}{\sqrt{24}}\right)$
in Eq.~(\ref{1}). In addition the splitting of bulk 27--plets can give rise to
a set of $M_{l}$ and $\overline{M}_l$ supermultiplets with opposite quantum numbers.

The presence of exotic matter in the $E_6$ inspired SUSY models generically
leads to rapid proton decay and non--diagonal flavour transitions at low energies.
To suppress flavour changing processes as well as the most dangerous baryon and
lepton number violating operators one can impose a single discrete $\tilde{Z}^{H}_2$
symmetry. All states from complete $27_i$--plets are odd whereas all
supermultiplets $M_{l}$ are even under this $\tilde{Z}^{H}_2$ symmetry.
Because $M_{l}$ can be used for the breakdown of gauge symmetry this set of
supermultiplets should contain $H_u$, $H_d$, $S$ and $N^c_H$. Since superfield
$N^c_H$ has the same $U(1)_{\psi}$ and $U(1)_{\chi}$ charges as the right--handed
neutrino the large vacuum expectation values (VEVs) of $N^c_H$ and $\overline{N}_H^c$
break $U(1)_{\psi}\times U(1)_{\chi}$ down to $U(1)_{N}$ generating large
Majorana masses for the right--handed neutrinos. Only in this $E_6$ inspired
$U(1)$ extension of the MSSM, i.e. in the so--called Exceptional
Supersymmetric Standard Model (E$_6$SSM)\cite{King:2005jy,King:2005my},
the right--handed neutrinos may be superheavy, shedding light on the origin of the
mass hierarchy in the lepton sector and providing a mechanism for the generation
of lepton and baryon asymmetry of the universe \cite{King:2008qb}.
Different phenomenological implications of the several variants of the E$_6$SSM were considered in
Refs.~\citenum{Nevzorov:2012hs,King:2005jy,King:2005my,King:2008qb,King:2006rh,King:2007uj,Athron:2010zz,Hall:2010ix,Hall:2010ny,Nevzorov:2013tta,Athron:2014pua,Nevzorov:2014sha,Nevzorov:2015iya,King:2016wep,Nevzorov:2017gir,Nevzorov:2013ixa}.
In particular, the renormalisation group (RG) flow of the gauge and Yukawa
couplings as well as the theoretical upper bound on the lightest Higgs boson
mass were examined in the vicinity of the quasi--fixed point \cite{Nevzorov:2013ixa}
that appears as a result of the intersection of the invariant and
quasi--fixed lines \cite{Nevzorov:2001vj,Nevzorov:2002ub}.
Within the constrained version of the E$_6$SSM and its modifications
the particle spectrum, the corresponding collider signatures and the
implications for dark matter were analysed in Refs.~\citenum{Athron:2009ue,Athron:2009bs,Athron:2011wu,Athron:2012sq,Athron:2015vxg,Athron:2016gor}.
Here we assume that $U(1)_{\psi}\times U(1)_{\chi}$ symmetry is broken down to
$U(1)_{N}\times Z_{2}^{M}$, where $Z_{2}^{M}=(-1)^{3(B-L)}$ is a matter parity.
This can occur because $Z_{2}^{M}$ is a discrete subgroup of $U(1)_{\psi}$ and
$U(1)_{\chi}$.

In the simplest case the set of the $Z^{H}_2$--even supermultiplets $M_{l}$
should also include a lepton $SU(2)_W$ doublet $L_4$ to allow the lightest
exotic quarks to decay \cite{Nevzorov:2012hs}. The supermultiplets $\overline{M}_l$
can be either even or odd under the $\tilde{Z}^{H}_2$ symmetry. The simplest scenario
imply that $\overline{S}$, $\overline{H}_u$ and $\overline{H}_d$ are odd whereas
$\overline{L}_4$ is even under $\tilde{Z}^{H}_2$. It is expected that the
$\tilde{Z}^{H}_2$-odd supermultiplets $\overline{S}$, $\overline{H}_u$ and $\overline{H}_d$
get combined with the superposition of the appropriate components from $27_i$
forming vectorlike states with masses of order of $M_X$. At the same time the
supermultiplets $L_4$ and $\overline{L}_4$ should form TeV scale vectorlike states
to render the lightest exotic quarks unstable.
The most general renormalisable superpotential which is allowed by the
$\tilde{Z}^{H}_2$, $Z_{2}^{M}$ and $SU(3)\times SU(2)_W\times U(1)_{Y}\times U(1)_{N}$
symmetries can be written as
\begin{equation}
\begin{array}{c}
W = \lambda S (H_u H_d) + \lambda_{\alpha\beta} S (H^d_{\alpha} H^u_{\beta})
+ \kappa_{ij} S (D_{i} \overline{D}_{j}) + \tilde{f}_{\alpha\beta} S_{\alpha} (H^d_{\beta} H_u) \\[1mm]
+ f_{\alpha\beta} S_{\alpha} (H_d H^u_{\beta})
+ h^E_{i\alpha} e^c_{i} (H^d_{\alpha} L_4) + \mu_L L_4\overline{L}_4 + W_N\\[1mm]
+ g^D_{ij} (Q_i L_4) \overline{D}_j + W_{\rm MSSM}(\mu=0)\,,
\end{array}
\label{3}
\end{equation}
where $\alpha,\beta=1,2$ and $i,j=1,2,3$ while
$W_{\rm MSSM}(\mu=0)$ is the MSSM superpotential with the bilinear mass parameter
$\mu$ set to zero and
\begin{equation}
W_N =  \frac{1}{2} M_{ij} N_i^c N_j^c + \tilde{h}_{ij} N_i^c (H_u L_j) +
h_{i\alpha}  N_i^c (H^u_{\alpha} L_4)\,.
\label{4}
\end{equation}
In Eqs.~(\ref{3}) and (\ref{4}) $e_i^c$ and $N^c_i$ are the right-handed charged
leptons and neutrinos whereas $Q_i$ and $L_j$ are the left-handed quark and
lepton doublets respectively.
The second last term in Eq.~(\ref{3}) ensures that the lightest exotic quarks
decay within a reasonable time when the couplings $g^D_{ij}$ are sufficiently large
and the components of the supermultiplets $L_4$ and $\overline{L}_4$ have masses
of the order of a few TeV. Since in this case extra matter beyond the MSSM fill
in complete $SU(5)$ representations the gauge coupling unification in the SUSY model
under consideration can be achieved for any phenomenologically acceptable value
of $\alpha_3(M_Z)$, consistent with its central measured low energy value
\cite{King:2007uj, Nevzorov:2012hs}.

The $Z^{H}_2$--even supermultiplets $H_u$, $H_d$ and $S$ gain
non--zero VEVs, i.e. $\langle H_d \rangle = v_1/\sqrt{2}$,
$\langle H_u \rangle = v_2/\sqrt{2}$ and $\langle S \rangle = s/\sqrt{2}$,
which are much smaller than the VEVs of $N^c_H$ and $\overline{N}_H^c$.
In phenomenologically viable scenarios the SM singlet superfield $S$ has
to acquire VEV which is much larger than $1\,\mbox{TeV}$ breaking
$U(1)_{N}$ gauge symmetry and inducing sufficiently large masses of $Z'$
boson and exotic fermion states. The neutral components of $H_u$ and $H_d$
develop VEVs, so that $v=\sqrt{v_1^2+v_2^2}\simeq 246\,\mbox{GeV}$.
These VEVs trigger the breakdown of the $SU(2)_{W}\times U(1)_{Y}$ symmetry
down to $U(1)_{em}$ associated with electromagnetism and give rise to
the masses of ordinary quarks and leptons.

In the framework of the E$_6$SSM the Higgs sector was explored in
Ref.~\citenum{King:2005jy}. When CP-invariance is preserved the Higgs spectrum
contains three CP-even, one CP-odd and two charged states. The SM singlet
dominated CP-even state and the $Z'$ gauge boson are almost degenerate.
If $\lambda < g'_1$, where $g'_1$ is the $U(1)_{N}$ gauge coupling,
the SM singlet dominated Higgs boson is the heaviest CP-even state.
In this case the rest of the Higgs spectrum is basically indistinguishable
from the one in the MSSM. When $\lambda\gtrsim g'_1$ the Higgs spectrum
has a very hierarchical structure, which is similar to the one
in the NMSSM with the approximate PQ symmetry \cite{Miller:2003ay}.
As a consequence the mass matrix of the CP--even Higgs sector can be
diagonalised using the perturbation theory \cite{Kovalenko:1998dc,Nevzorov:2000uv,Nevzorov:2001um}.
If $\lambda\gtrsim g'_1$ the MSSM--like CP-even, CP-odd and charged states
have almost the same masses and lie beyond the TeV range.

For the analysis of the phenomenological implications of the SUSY
models discussed above it is convenient to introduce the $Z_{2}^{E}$
symmetry, which can be defined such that $\tilde{Z}^{H}_2 = Z_{2}^{M}\times Z_{2}^{E}$.
The supermultiplets $S_{\alpha}$, $H^{u}_{\alpha}$, $H^{d}_{\alpha}$,
$D_i$, $\bar{D}_i$, $L_4$ and $\overline{L}_4$ are odd under the
$Z_{2}^{E}$ symmetry. The components of all other supermultiplets are
$Z_{2}^{E}$ even. Because the Lagrangian is invariant under both
$Z_{2}^{M}$ and $\tilde{Z}^{H}_2$ symmetries, the $Z_{2}^{E}$ symmetry
is also conserved. This implies that in collider experiments the exotic
particles, which are odd under the $Z_{2}^{E}$ symmetry, can only be created
in pairs and the lightest exotic state has to be absolutely stable.
Using the method proposed in Ref.~\citenum{Hesselbach:2007te} it was argued
that the masses of the lightest exotic fermions, which are predominantly
linear superpositions of the fermion components of the superfields
$S_{\alpha}$, do not exceed $60-65\,\mbox{GeV}$ \cite{Hall:2010ix}.
Thus these states tend to be the lightest exotic particles in the
spectrum. Moreover the lightest exotic fermion is also the lightest SUSY
particle (LSP). Although the couplings of the corresponding states to the
SM gauge bosons and fermions are quite small the lightest exotic state
could account for all or some of the observed cold dark matter density
if it had a mass close to half the $Z$ mass. However in this case
the SM--like Higgs boson would decay almost 100\% of the time into
the fermion components of $S_{\alpha}$.  All other branching ratios would be
strongly suppressed. Basically such scenario has been already ruled
out by the LHC experiments. On the other hand if the lightest exotic
fermions are substantially lighter than $M_Z$ the annihilation cross
section for $\mbox{LSP}+\mbox{LSP} \to \mbox{SM particles}$ becomes too small
leading to a relic density that is much larger than its measured value.

The simplest phenomenologically viable scenarios imply that the fermion components
of $S_{\alpha}$ are significantly lighter than $1\,\mbox{eV}$\footnote{The presence of
very light neutral fermions in the particle spectrum might have interesting
implications for neutrino physics \cite{Frere:1996gb}.}.
In this scenario the lightest SUSY particles form hot dark matter in the Universe.
When the masses of the fermion components of $S_{\alpha}$ are considerably smaller
than $1\,\mbox{eV}$ these states give only a very minor contribution to
the dark matter density. At the same time the invariance of the Lagrangian under
the $Z_{2}^{M}$ symmetry ensures that the $R$-parity is also conserved and
the lightest ordinary neutralino is stable. In this case the lightest ordinary
neutralino may account for all or some of the observed cold dark matter density.

The scenarios discussed above are realised if
$\tilde{f}_{\alpha\beta}\sim f_{\alpha\beta}\lesssim 10^{-6}$.
When the Yukawa couplings of the superfields $S_{\alpha}$ are very small
the terms $\tilde{f}_{\alpha\beta} S_{\alpha} (H^d_{\beta} H_u)$ and
$f_{\alpha\beta} S_{\alpha} (H_d H^u_{\beta})$ in the superpotential (\ref{3})
can be ignored. In this limit the low--energy effective Lagrangian
possesses an approximate global $U(1)_E$ symmetry below the scale $M_1$
where $M_1$ is the mass of the lightest right--handed neutrinos.
The $U(1)_E$ charges of the exotic matter fields are summarised
in Table~\ref{tab1}. Both $U(1)_{B-L}$ and $U(1)_E$ symmetries are
explicitly broken because of the interactions of matter supermultiplets
with $N_i^c$ in $W_N$. As a consequence the decays of the lightest
right--handed neutrino/sneutrino induce simultaneously $U(1)_E$ and
$U(1)_{B-L}$ asymmetries. These asymmetries would not be washed out
in the limit $\tilde{f}_{\alpha\beta}, f_{\alpha\beta} \to 0$. Moreover the
sufficiently small values of the $U(1)_E$ violating Yukawa couplings,
i.e $\tilde{f}_{\alpha\beta}, f_{\alpha\beta} \lesssim 10^{-7}$, should not
erase the induced $U(1)_E$ asymmetry \cite{Campbell:1990fa,Davidson:1996hs,Mohapatra:2015fua}.
The non-zero values of $\tilde{f}_{\alpha\beta}$ and $f_{\alpha\beta}$ break
the $U(1)_E$ symmetry and the lightest exotic state that carries the $U(1)_E$ charge
becomes unstable. It decays into the fermion components of $S_{\alpha}$ so that
the generated $U(1)_E$ asymmetry gets converted into the hot dark matter
density.

\begin{table}
\tbl{The $U(1)_E$ charges of exotic matter supermultiplets.}
{\begin{tabular}{@{}cccccc@{}}
\toprule
\qquad $H^{u}_{\alpha}$ & \qquad $H^{d}_{\alpha}$ & \qquad $D_i$ & \qquad $\overline{D}_i$ & \qquad $L_4$
& \qquad $\overline{L}_4$ \\
\colrule
\qquad +1 \qquad  & \qquad -1 \qquad &\qquad +1 \qquad &\qquad -1 \qquad &\qquad +1 \qquad &\qquad -1 \qquad   \\
\botrule
\end{tabular}
}
\label{tab1}
\end{table}

\section{Generation of baryon asymmetry}
A potential drawback of supersymmetric thermal leptogenesis is the
lower bound on $M_1$. Indeed, it was shown that the appropriate amount
of the baryon asymmetry in the SM and MSSM can be induced only
if $M_1$ is larger than $10^9\,\mbox{GeV}$ \cite{Davidson:2002qv,Hamaguchi:2001gw}.
In the framework of supergravity this lower bound on $M_1$
leads to the gravitino problem \cite{Khlopov:1984pf,Ellis:1984eq}.
After inflation the universe thermalizes with a reheat
temperature $T_R$. If $T_R> M_1$, the right-handed neutrinos are
produced by thermal scattering and thermal leptogenesis could take place.
At the same time when $T_R\gtrsim 10^9\,\mbox{GeV}$ such a high
reheating temperature results in an overproduction of gravitinos
which tend to decay during or after Big Bang Nucleosynthesis (BBN)
destroying the agreement between the predicted and observed light
element abundances. It was argued that the gravitino density becomes
low enough if $T_R\lesssim 10^{6-7}\,\mbox{GeV}$
\cite{Khlopov:1993ye,Kawasaki:2004qu,Kohri:2005wn}.

In order to avoid the gravitino problem we fix $M_1\simeq 10^6\,\mbox{GeV}$.
We also assume that two other right-handed neutrino states have masses
$M_{2,3}\sim 10^{6-7}\,\mbox{GeV}$. For so low $M_i$ the absolute values of
the Yukawa couplings $|\tilde{h}_{ij}|$ should be rather small to reproduce
the left--handed neutrino mass scale $m_{\nu}\lesssim 0.1\,\mbox{eV}$,
i.e. $|\tilde{h}_{ij}|^2\ll 10^{-8}$. So small Yukawa couplings can be
ignored in the leading approximation. Then only the new channels of the decays
of the lightest right--handed neutrino $N_1$ and its superpartner $\widetilde{N}_1$,
i.e.
\begin{equation}
N_1\to L_4 + H^u_{\alpha},\quad N_1\to \widetilde{L}_4+\widetilde{H}^u_{\alpha},\quad
\widetilde{N}^{*}_1\to L_4 + \widetilde{H}^{u}_{\alpha},\quad \widetilde{N}_1\to \widetilde{L}_4+ H^u_{\alpha},
\label{5}
\end{equation}
give rise to the generation of lepton asymmetry. This process is controlled by the
CP (decay) asymmetries associated with the decays of $N_1$, i.e.
\begin{equation}
\varepsilon^{\alpha}_{1,\,\ell_4}=\frac{\Gamma^{\alpha}_{N_1 \ell_4}-\Gamma^{\alpha}_{N_1 \bar{\ell}_4}}
{\sum_{\beta} \left(\Gamma^{\beta}_{N_1 \ell_4}+\Gamma^{\beta}_{N_1 \bar{\ell}_4}\right)}\,,\qquad
\varepsilon^{\alpha}_{1,\,\widetilde{\ell}_4}=\frac{\Gamma^{\alpha}_{N_1 \widetilde{\ell}_{4}}
-\Gamma^{\alpha}_{N_1 \widetilde{\ell}^{*}_{4}}}
{\sum_{\beta} \left(\Gamma^{\beta}_{N_1 \widetilde{\ell}_{4}}+
\Gamma^{\beta}_{N_1 \widetilde{\ell}^{*}_{4}}\right)}\,,
\label{6}
\end{equation}
and $\widetilde{N}_1$, i.e.
\begin{equation}
\varepsilon^{\alpha}_{\widetilde{1},\,\ell_4}=\frac{\Gamma^{\alpha}_{\widetilde{N}_1^{*}\ell_{4}}
-\Gamma^{\alpha}_{\widetilde{N}_1 \bar{\ell}_{4}}}
{\sum_{\beta} \left(\Gamma^{\beta}_{\widetilde{N}_1^{*} \ell_{4}}+
\Gamma^{\beta}_{\widetilde{N}_1 \bar{\ell}_{4}}\right)}\,,\qquad
\varepsilon^{\alpha}_{\widetilde{1},\,\widetilde{\ell}_4}=\frac{\Gamma^{\alpha}_{\widetilde{N}_1\widetilde{\ell}_{4}}
-\Gamma^{\alpha}_{\widetilde{N}_1^{*} \widetilde{\ell}^{*}_{4}}}
{\sum_{\beta} \left(\Gamma^{\beta}_{\widetilde{N}_1 \widetilde{\ell}_{4}}+
\Gamma^{\beta}_{\widetilde{N}_1^{*} \widetilde{\ell}^{*}_{4}}\right)}\,.
\label{7}
\end{equation}
In Eqs.~(\ref{6}) and (\ref{7}) the superscripts $\alpha$ and $\beta$ represent
the components of the supermultiplets $H^{u}_{\alpha}$ and $H^{u}_{\beta}$ in the final state.
At the tree level the partial decay widths associated with the new channels
(\ref{5}) are given by
\begin{equation}
\Gamma^{\alpha}_{N_1 \ell_4}+\Gamma^{\alpha}_{N_1 \bar{\ell}_4}=\Gamma^{\alpha}_{N_1 \widetilde{\ell}_4}+\Gamma^{\alpha}_{N_1 \widetilde{\ell}^{*}_4}=
\Gamma^{\alpha}_{\widetilde{N}_1^{*}\ell_4}=\Gamma^{\alpha}_{\widetilde{N}_1 \bar{\ell}_4}=\Gamma^{\alpha}_{\widetilde{N}_1 \widetilde{\ell}_4}=
\Gamma^{\alpha}_{\widetilde{N}_1^{*} \widetilde{\ell}_4^{*}}=\frac{|h_{1\alpha}|^2}{8\pi}M_1
\label{8}
\end{equation}
and all decay asymmetries (\ref{6}) and (\ref{7}) vanish.

The non--zero values of the CP asymmetries arise after the
inclusion of one--loop vertex and self--energy corrections to the decay
amplitudes of $N_1$ and $\widetilde{N}_1$. In this context
it is worth noting that the supermultiplets $H^{u}_{\alpha}$ can be
redefined so that only one doublet $H^{u}_{1}$ interacts with $L_4$
and $N^c_1$. Therefore without loss of generality $h_{12}$ in $W_N$
may be set to zero. In this limit
$\varepsilon^{2}_{1,\,\ell_4}=\varepsilon^{2}_{1,\,\widetilde{\ell}_4}=
\varepsilon^{2}_{\widetilde{1},\,\ell_4}=
\varepsilon^{2}_{\widetilde{1},\,\widetilde{\ell}_4}=0$.
When SUSY breaking scale is negligibly small as compared with $M_1$,
$h_{j1}=|h_{j1}| e^{i\varphi_{j1}}$ and $M_j$ are real the non--zero
asymmetries are given by
\begin{equation}
\varepsilon^{1}_{1,\,\ell_4}=\varepsilon^{1}_{1,\,\widetilde{\ell}_4}=
\varepsilon^{1}_{\widetilde{1},\,\ell_4}=\varepsilon^{1}_{\widetilde{1},\,\widetilde{\ell}_4}=
\frac{1}{8\pi}\Biggl[\sum_{j=2,3} |h_{j1}|^2 f\left(\frac{M^2_j}{M_1^2}\right) \sin 2\Delta\varphi_{j1}\Biggr]\,,
\label{12}
\end{equation}
where $\Delta\varphi_{j1}=\varphi_{j1}-\varphi_{11}$ and
\begin{equation}
f(z)=f^{V}(z)+f^{S}(z)\,,\qquad f^S(z)=\dfrac{2\sqrt{z}}{1-z}\,,\qquad f^V(z)=-\sqrt{z}\,\ln\left(\dfrac{1+z}{z}\right)\,.
\label{11}
\end{equation}
Because the Yukawa couplings of the superfields $N^c_i$ to the
supermultiplets $H^{u}_{\alpha}$ and $L_4$ violate both $U(1)_E$ and $U(1)_{B-L}$
the decay channels of the lightest right--handed neutrino/sneutrino (\ref{5})
induce simultaneously $U(1)_{B-L}$ and $U(1)_E$ asymmetries. These asymmetries
are determined by the same set of the CP asymmetries (\ref{12}).

\begin{figure}[h]
\begin{center}
\includegraphics[width=2.5in]{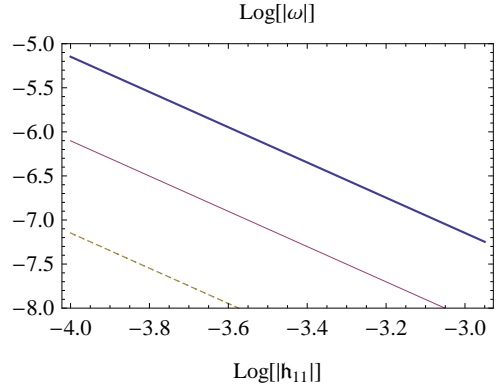}
\end{center}
\caption{Logarithm (base 10) of the absolute value of $\omega=\varepsilon^{1}_{1,\,\ell_4} \eta$
as a function of logarithm (base 10) of $|h_{11}|$ for $h_{i2}=h_{31}=0$, $\Delta\varphi_{21}=\pi/4$
and $M_2=10\cdot M_1$. The thick, solid and dashed lines correspond to $|h_{21}|=0.3$, $|h_{21}|=0.1$ and $|h_{21}|=0.03$ respectively.}
\label{fig1}
\end{figure}

The evolution of the $U(1)_{B-L}$ and $U(1)_E$ asymmetries are
described by the system of Boltzmann equations. The generated baryon
asymmetry can be estimated as follows
\begin{equation}
Y_{\Delta B}\sim 10^{-3} \varepsilon^{1}_{1,\,\ell_4} \eta \,,
\label{13}
\end{equation}
where $Y_{\Delta B}$ is the baryon asymmetry relative
to the entropy density, i.e.
\begin{equation}
Y_{\Delta B}=\dfrac{n_B-n_{\bar{B}}}{s}\biggl|_0=(8.75\pm 0.23)\times 10^{-11}\,.
\label{14}
\end{equation}
In Eq.~(\ref{13}) $\eta$ is an efficiency factor. It varies from 0 to 1.
In the strong washout scenario $\eta$ is given by
\begin{equation}
\eta \simeq H(T=M_1)/\Gamma_{1}\,,
\label{15}
\end{equation}
where $H$ is the Hubble expansion rate
\begin{equation}
H=1.66 g_{*}^{1/2}\dfrac{T^2}{M_{Pl}}\,,
\label{16}
\end{equation}
$g_{*}=n_b+\dfrac{7}{8}\,n_f$ is the number of relativistic degrees
of freedom and
\begin{equation}
\Gamma_1 = \Gamma^{1}_{N_1 \ell_4}+\Gamma^{1}_{N_1 \bar{\ell}_4}=\dfrac{|h_{11}|^2}{8\pi}\,M_1\,.
\label{17}
\end{equation}

As follows from Eq.~(\ref{12}) the values of the CP asymmetries are
determined by the CP--violating phases $\Delta\varphi_{j1}$ and the
absolute values of the Yukawa couplings $|h_{21}|$ and $|h_{31}|$ but
do not depend on $|h_{11}|$. To simplify our analysis we fix
$|h_{31}|=0$ and $(M_2/M_1)=10$. At the same time the efficiency
factor $\eta$ is set by the lightest right--handed neutrino mass $M_1$
and $|h_{11}|$. We restrict our consideration here by the values of
$|h_{11}|^2 \gg |\tilde{h}_{ik}|^2$, i.e. $|h_{11}|^2 \gtrsim 10^{-8}$.
For $\Delta\varphi_{21}=\pi/4$ we find
\begin{equation}
\log |\eta|\simeq -2\log |h_{11}| - 10.2\,,\qquad\qquad
\log |\omega|\simeq - 2\log |h_{11}| + 2\log |h_{21}| - 12.1\,,
\label{18}
\end{equation}
where $\omega=\varepsilon^{1}_{1,\,\ell_4} \eta$. Eq.~(\ref{18})
indicates that $\eta$ varies from $10^{-2}$ to $10^{-4}$ when $|h_{11}|$
increases from $10^{-4}$ to $10^{-3}$. The dependence of $|\omega|$,
that determines the generated baryon asymmetry (\ref{13}),
on $|h_{21}|$ and $|h_{11}|$ is explored in Fig.~\ref{fig1}.
This figure illustrates that for $\Delta\varphi_{21}=\pi/4$ and
$|h_{21}|\sim 0.1$ the phenomenologically acceptable baryon
density, corresponding to $\omega \sim 10^{-7}-10^{-6}$, can
be obtained if $|h_{11}|$ varies between $10^{-4}$ and $10^{-3}$.
If $\tilde{f}_{\alpha\beta}, f_{\alpha\beta} \lesssim 10^{-7}$
the induced dark matter and baryon number densities should be
of the same order of magnitude.

\section{Exotic Higgs decays}
As it was mentioned before the lightest and second lightest exotic states
($\chi^0_1$ and $\chi^0_2$) are mostly linear superpositions of the
fermion components of the superfields $S_{\alpha}$. In the simplest
phenomenologically viable scenarios
$\chi^0_1$ should have mass $m_{\chi_1}\ll 1\,\mbox{eV}$.
At the same time $\chi^0_2$
can be considerably heavier if some of the Yukawa couplings
$\tilde{f}_{\alpha\beta}$ and $f_{\alpha\beta}$ are much larger than
$10^{-6}-10^{-5}$. Although $\chi^0_1$ and $\chi^0_2$ tend to be rather
light their couplings to the $Z$--boson and other SM particles
can be negligibly small because these states are predominantly the
fermion components of the SM singlet superfields
$S_{\alpha}$ \cite{Nevzorov:2013tta}. As a result
any possible signal, which $\chi^0_1$ and $\chi^0_2$ could give rise to
at former and present collider experiments, would be extremely suppressed and
such states could escape their experimental detection.

The couplings of the lightest Higgs boson $h_1$ to $\chi^0_1$
and $\chi^0_2$ are determined by their masses \cite{Hall:2010ix}. Since
$\chi^0_1$ is extremely light it does not affect Higgs phenomenology.
The coupling of the SM--like Higgs state $h_1$ to the second lightest exotic
particle $X^{h}_{22}\simeq |m_{\chi_{2}}|/v$ \cite{Hall:2010ix}.
This coupling gives rise to the decays of $h_1$ into $\chi^0_2$ pairs
with partial width given by
\begin{equation}
\Gamma(h_1\to\chi^0_{2}\chi^0_{2})=\frac{(X^{h}_{22})^2 m_{h_1}}{4\pi}
\left(1-4\frac{|m_{\chi_{2}}|^2}{m^2_{h_1}}\right)^{3/2}\,,
\label{19}
\end{equation}
where $m_{h_1}$ is the lightest Higgs boson mass. From Eq.~(\ref{19})
it follows that the partial decay width of the non--standard Higgs decays
depend rather strongly on $m_{\chi_{2}}$. The branching ratio of
$h_1\to \chi^0_2 \chi^0_2$, can be substantial if the second lightest exotic
fermion has a mass of order of the $b$--quark mass $m_b$. To avoid the
suppression of the branching ratios for Higgs decays into SM particles
we restrict our consideration to the GeV scale masses of the second
lightest exotic particle.

After being produced $\chi^0_2$ sequentially decay into $\chi^0_1$
and fermion--antifermion pair via virtual $Z$. Thus the exotic decays
of the SM--like Higgs discussed above results in two fermion--antifermion
pairs and missing energy in the final state. Nevertheless due to the
small coupling of the lightest and second lightest exotic fermions
to the $Z$--boson $\chi^0_2$ tends to live longer than $10^{-8}\,\mbox{sec}$.
Therefore it typically decays outside the detectors and can not be
observed at the LHC directly. As a consequence the decay channel
$h_1\to\chi^0_2\chi^0_2$ normally gives rise to an invisible branching
ratio of $h_1$. If the second lightest exotic fermion is very long-lived
then $\chi^0_2$ may decay during or after Big Bang Nucleosynthesis (BBN)
destroying the agreement between the predicted and observed light element
abundances. To preserve the success of the BBN, the lifetime
$\tau_{\chi_{2}}$ of $\chi^0_2$ should not be longer than $1\,\mbox{sec}$.
Because $\tau_{\chi_{2}}\sim 1/(m_{\chi_{2}}^5)$ this requirement basically
rules out too light $\chi^0_2$. Indeed, it is somewhat problematic
to satisfy this restriction for $m_{\chi_{2}}\lesssim 100\,\mbox{MeV}$.

The numerical analysis indicates that the branching ratio associated with
the decays $h_1\to \chi^0_2\chi^0_2$ can vary from $0.2\%$ to $20\%$
when $m_{\chi_{2}}$ changes from $0.3\,\mbox{GeV}$ to
$2.7\,\mbox{GeV}$ \cite{Nevzorov:2013tta}.
When $\chi^0_2$ is lighter than 0.5 GeV the corresponding
branching ratio can be as small as $10^{-3}-10^{-4}$.



\begin{thebibliography}{10}

\bibitem{Ellis:1990wk}
J.~R. Ellis, S.~Kelley and D.~V. Nanopoulos,
Probing the desert using gauge coupling unification,
{\em Phys. Lett. B} {\bf 260}, 131 (1991).

\bibitem{Langacker:1991an}
P.~Langacker and M.~Luo,
Implications of precision electroweak experiments for $M_t$, $\rho_{0}$, $\sin^2\theta_W$
and grand unification,
{\em Phys. Rev. D} {\bf 44}, 817 (1991).

\bibitem{Amaldi:1991cn}
U.~Amaldi, W.~de Boer and H.~Furstenau,
Comparison of grand unified theories with electroweak and strong coupling constants measured at LEP,
{\em Phys. Lett. B} {\bf 260}, 447 (1991).

\bibitem{Anselmo:1991uu}
F.~Anselmo, L.~Cifarelli, A.~Peterman and A.~Zichichi,
The Effective experimental constraints on M (susy) and M (gut),
{\em Nuovo Cim. A} {\bf 104}, 1817 (1991).

\bibitem{5}
M.~B. Green, J.~H. Schwarz, E.~Witten, {\em Superstring Theory} (Cambridge Univ. Press, Cambridge, 1987).

\bibitem{Nevzorov:2012hs}
R.~Nevzorov,
$E_6$ inspired supersymmetric models with exact custodial symmetry,
{\em Phys. Rev. D} {\bf 87}, 015029  (2013).

\bibitem{King:2005jy}
S.~F. King, S.~Moretti and R.~Nevzorov,
Theory and phenomenology of an exceptional supersymmetric standard model,
{\em Phys. Rev. D} {\bf 73}, 035009  (2006).

\bibitem{King:2005my}
S.~F. King, S.~Moretti and R.~Nevzorov,
Exceptional supersymmetric standard model,
{\em Phys. Lett. B} {\bf 634}, 278 (2006).

\bibitem{King:2008qb}
S.~F. King, R.~Luo, D.~J. Miller and R.~Nevzorov,
Leptogenesis in the Exceptional Supersymmetric Standard Model: Flavour dependent lepton asymmetries,
{\em JHEP} {\bf 0812}, 042 (2008).

\bibitem{King:2006rh}
S.~F. King, S.~Moretti and R.~Nevzorov,
E$_6$SSM,
{\em AIP Conf. Proc.} {\bf 881}, 138 (2007).

\bibitem{King:2007uj}
S.~F. King, S.~Moretti and R.~Nevzorov,
Gauge coupling unification in the exceptional supersymmetric standard model,
{\em Phys. Lett. B} {\bf 650}, 57 (2007).

\bibitem{Athron:2010zz}
P.~Athron, J.~P. Hall, R.~Howl, S.~F. King, D.~J. Miller, S.~Moretti and R.~Nevzorov,
Aspects of the exceptional supersymmetric standard model,
{\em Nucl. Phys. Proc. Suppl.}  {\bf 200-202}, 120 (2010).

\bibitem{Hall:2010ix}
J.~P. Hall, S.~F. King, R.~Nevzorov, S.~Pakvasa and M.~Sher,
Novel Higgs Decays and Dark Matter in the E$_6$SSM,
{\em Phys. Rev. D} {\bf 83}, 075013 (2011).

\bibitem{Hall:2010ny}
J.~P. Hall, S.~F. King, R.~Nevzorov, S.~Pakvasa and M.~Sher,
Nonstandard Higgs decays in the E$_6$SSM,
{\em PoS QFTHEP} {\bf 2010}, 069 (2010).

\bibitem{Nevzorov:2013tta}
R.~Nevzorov and S.~Pakvasa,
Exotic Higgs decays in the $E_6$ inspired SUSY models,
{\em Phys. Lett. B} {\bf 728}, 210 (2014).

\bibitem{Athron:2014pua}
P.~Athron, M.~Mühlleitner, R.~Nevzorov and A.~G. Williams,
Non-Standard Higgs Decays in U(1) Extensions of the MSSM,
{\em JHEP} {\bf 1501}, 153 (2015).

\bibitem{Nevzorov:2014sha}
R.~Nevzorov and S.~Pakvasa,
Nonstandard Higgs decays in the E$_6$ inspired SUSY models,
{\em Nucl. Part. Phys. Proc.}  {\bf 273-275}, 690 (2016).

\bibitem{Nevzorov:2015iya}
R.~Nevzorov,
LHC Signatures and Cosmological Implications of the E$_6$ Inspired SUSY Models,
{\em PoS EPS} {\bf -HEP2015}, 381 (2015).

\bibitem{King:2016wep}
S.~F. King and R.~Nevzorov,
750 GeV Diphoton Resonance from Singlets in an Exceptional Supersymmetric Standard Model,
{\em JHEP} {\bf 1603}, 139 (2016).

\bibitem{Nevzorov:2017gir}
R.~Nevzorov,
Leptogenesis as an origin of hot dark matter and baryon asymmetry in the $E_6$ inspired SUSY models,
{\em Phys. Lett. B} {\bf 779}, 223 (2018).

\bibitem{Nevzorov:2013ixa}
R.~Nevzorov,
Quasifixed point scenarios and the Higgs mass in the E6 inspired supersymmetric models,
{\em Phys. Rev. D} {\bf 89}, 055010 (2014).

\bibitem{Nevzorov:2001vj}
R.~B. Nevzorov and M.~A. Trusov,
Infrared quasifixed solutions in the NMSSM,
{\em Phys. Atom. Nucl.}  {\bf 64}, 1299 (2001).

\bibitem{Nevzorov:2002ub}
R.~B. Nevzorov and M.~A. Trusov,
Quasifixed point scenario in the modified NMSSM,
{\em Phys. Atom. Nucl.}  {\bf 65}, 335 (2002).


\bibitem{Athron:2009ue}
P.~Athron, S.~F. King, D.~J. Miller, S.~Moretti and R.~Nevzorov,
Predictions of the Constrained Exceptional Supersymmetric Standard Model,
{\em Phys. Lett. B} {\bf 681}, 448 (2009).

\bibitem{Athron:2009bs}
P.~Athron, S.~F. King, D.~J. Miller, S.~Moretti and R.~Nevzorov,
The Constrained Exceptional Supersymmetric Standard Model,
{\em Phys. Rev. D} {\bf 80}, 035009 (2009).

\bibitem{Athron:2011wu}
P.~Athron, S.~F. King, D.~J. Miller, S.~Moretti and R.~Nevzorov,
LHC Signatures of the Constrained Exceptional Supersymmetric Standard Model,
{\em Phys. Rev. D} {\bf 84}, 055006 (2011).

\bibitem{Athron:2012sq}
P.~Athron, S.~F. King, D.~J. Miller, S.~Moretti and R.~Nevzorov,
Constrained Exceptional Supersymmetric Standard Model with a Higgs Near 125 GeV,
{\em Phys. Rev. D} {\bf 86}, 095003 (2012).

\bibitem{Athron:2015vxg}
P.~Athron, D.~Harries, R.~Nevzorov and A.~G. Williams,
$E_6$ Inspired SUSY benchmarks, dark matter relic density and a 125 GeV Higgs,
{\em Phys. Lett. B} {\bf 760}, 19 (2016).

\bibitem{Athron:2016gor}
P.~Athron, D.~Harries, R.~Nevzorov and A.~G. Williams,
Dark matter in a constrained $E_{6}$ inspired SUSY model,
{\em JHEP} {\bf 1612}, 128 (2016).

\bibitem{Miller:2003ay}
D.~J. Miller, R.~Nevzorov and P.~M. Zerwas,
The Higgs sector of the next-to-minimal supersymmetric standard model,
{\em Nucl. Phys. B} {\bf 681}, 3 (2004).

\bibitem{Kovalenko:1998dc}
P.~A. Kovalenko, R.~Nevzorov and K.~A. Ter-Martirosian,
Masses of Higgs bosons in supersymmetric theories,
{\em Phys. Atom. Nucl.} {\bf 61}, 812 (1998).

\bibitem{Nevzorov:2000uv}
R.~Nevzorov and M.~A. Trusov,
Particle spectrum in the modified NMSSM in the strong Yukawa coupling limit,
{\em J. Exp. Theor. Phys.} {\bf 91}, 1079 (2000).

\bibitem{Nevzorov:2001um}
R.~Nevzorov, K.~A. Ter-Martirosyan and M.~A. Trusov,
Higgs bosons in the simplest SUSY models,
{\em Phys. Atom. Nucl.} {\bf 65}, 285 (2002).

\bibitem{Hesselbach:2007te}
S.~Hesselbach, D.~J. Miller, G.~Moortgat-Pick, R.~Nevzorov and M.~Trusov,
Theoretical upper bound on the mass of the LSP in the MNSSM,
{\em Phys. Lett. B} {\bf 662}, 199 (2008).

\bibitem{Frere:1996gb}
J.~M. Frere, R.~Nevzorov and M.~I. Vysotsky,
Stimulated neutrino conversion and bounds on neutrino magnetic moments,
{\em Phys. Lett. B} {\bf 394}, 127 (1997).

\bibitem{Campbell:1990fa}
B.~A. Campbell, S.~Davidson, J.~R. Ellis and K.~A. Olive,
Cosmological baryon asymmetry constraints on extensions of the standard model,
{\em Phys. Lett. B} {\bf 256}, 484 (1991).

\bibitem{Davidson:1996hs}
S.~Davidson and R.~Hempfling,
Protecting the baryon asymmetry in theories with R-parity violation,
{\em Phys. Lett. B} {\bf 391}, 287 (1997).

\bibitem{Mohapatra:2015fua}
R.~N. Mohapatra,
Supersymmetry and R-parity: an Overview,
{\em Phys. Scripta} {\bf 90}, 088004 (2015).

\bibitem{Davidson:2002qv}
S.~Davidson and A.~Ibarra,
A Lower bound on the right-handed neutrino mass from leptogenesis,
{\em Phys. Lett. B} {\bf 535}, 25 (2002).

\bibitem{Hamaguchi:2001gw}
K.~Hamaguchi, H.~Murayama and T.~Yanagida,
Leptogenesis from N dominated early universe,
{\em Phys. Rev. D} {\bf 65}, 043512 (2002).

\bibitem{Khlopov:1984pf}
M.~Y. Khlopov and A.~D. Linde,
Is It Easy to Save the Gravitino?,
{\em Phys. Lett. B} {\bf 138}, 265 (1984).

\bibitem{Ellis:1984eq}
J.~R. Ellis, J.~E. Kim and D.~V. Nanopoulos,
Cosmological Gravitino Regeneration and Decay,
{\em Phys. Lett. B} {\bf 145}, 181 (1984).

\bibitem{Khlopov:1993ye}
M.~Y. Khlopov, Y.~L. Levitan, E.~V. Sedelnikov and I.~M. Sobol,
Nonequilibrium cosmological nucleosynthesis of light elements:
Calculations by the Monte Carlo method,
{\em Phys. Atom. Nucl.} {\bf 57}, 1393 (1994).

\bibitem{Kawasaki:2004qu}
M.~Kawasaki, K.~Kohri and T.~Moroi,
Big-Bang nucleosynthesis and hadronic decay of long-lived massive particles,
{\em Phys. Rev. D} {\bf 71}, 083502 (2005).

\bibitem{Kohri:2005wn}
K.~Kohri, T.~Moroi and A.~Yotsuyanagi,
Big-bang nucleosynthesis with unstable gravitino and upper bound on the reheating temperature,
{\em Phys. Rev. D} {\bf 73}, 123511 (2006).

\end{thebibliography}
\end{document}